\documentclass{article}

\PassOptionsToPackage{numbers}{natbib}
\usepackage[final]{infocog_neurips_2022}

\usepackage[utf8]{inputenc} % allow utf-8 input
\usepackage[T1]{fontenc}    % use 8-bit T1 fonts
\usepackage{hyperref}       % hyperlinks
\usepackage{url}            % simple URL typesetting
\usepackage{booktabs}       % professional-quality tables
\usepackage{amsfonts}       % blackboard math symbols
\usepackage{nicefrac}       % compact symbols for 1/2, etc.
\usepackage{microtype}      % microtypography
\usepackage{xcolor}         % colors
\usepackage[pdftex]{graphicx} % figures

\title{Compression supports low-dimensional representations of behavior across neural circuits}

\author{%
Dale Zhou$^{1}$ \quad Jason Z. Kim$^{2}$ \quad Adam R. Pines$^{3}$ \quad Valerie J. Sydnor$^{1}$ \\ \textbf{David R. Roalf}$^{1}$ \quad \textbf{John A. Detre}$^{1}$ \quad \textbf{Ruben C. Gur}$^{1}$ \quad \textbf{Raquel E. Gur}$^{1}$ \\ \textbf{Theodore D. Satterthwaite}$^{1}$ \quad \textbf{Dani S. Bassett*}$^{1,4}$\\ $^{1}$University of Pennsylvania \quad $^{2}$Cornell University \quad $^{3}$Stanford University \quad $^{4}$Santa Fe Institute\\ \texttt{\{dalezhou,pinesa,valerie.sydnor,roalf,detre,gur,raquel,sattertt\}} \\ \texttt{@pennmedicine.upenn.edu}, \texttt{jk2557@cornell.edu}, \texttt{dsb@seas.upenn.edu}
}

\begin{document}
\maketitle
\begin{abstract}
Dimensionality reduction, a form of compression, can simplify representations of information to increase efficiency and reveal general patterns. Yet, this simplification also forfeits information, thereby reducing representational capacity. Hence, the brain may benefit from generating both compressed and uncompressed activity, and may do so in a heterogeneous manner across diverse neural circuits that represent low-level (sensory) or high-level (cognitive) stimuli. However, precisely how compression and representational capacity differ across the cortex remains unknown. Here we predict different levels of compression across regional circuits by using random walks on networks to model activity flow and to formulate rate-distortion functions, which are the basis of lossy compression. Using a large sample of youth ($n=1,040$), we test predictions in two ways: by measuring the dimensionality of spontaneous activity from sensorimotor to association cortex, and by assessing the representational capacity for 24 behaviors in neural circuits and 20 cognitive variables in recurrent neural networks. Our network theory of compression predicts the dimensionality of activity ($t=12.13, p<0.001$) and the representational capacity of biological ($r=0.53, p=0.016$) and artificial ($r=0.61, p<0.001$) networks. The model suggests how a basic form of compression is an emergent property of activity flow between distributed circuits that communicate with the rest of the network.
\end{abstract}

\section{Introduction}

Brain function is metabolically expensive \cite{attwell2001energy}. Neural processes can maximize efficiency by using lossy compression \cite{barlow1961possible, laughlin2001energy}, which compactly represents information with a given error \cite{Shannon1959CodingTF}. Compressed representations are efficiently stored, manipulated, and communicated \cite{eliasmith2012large}. By discarding redundant or irrelevant information, compressed representations can foreground key features and generalizable patterns \cite{sims2018efficient, mack2020ventromedial}.  Consequently, compression is a compelling model for cognitive functions that sacrifice the capacity to represent total information in order to efficiently segment and summarize relevant information. For example, working memory segments items into chunks \cite{brady2009compression, bates2019adaptive, hedayati2022model}; emotion processing summarizes interoceptive information \cite{barrett2017emotions}; and value estimation summarizes promising options for action \cite{lai2021policy}. The notion that compression is functionally important for brain and artificial networks \cite{srinivasan1982predictive, higgins2016beta, farrell2022gradient} is the \textit{cognitive compression hypothesis}.

A key challenge to the cognitive compression hypothesis is neural evidence. One way in which compression can be implemented is by \textit{efficient coding}, whereby the brain may reduce redundancy (e.g., activates fewer neurons) to maximize information capacity and economically use limited resources for behavior \cite{barlow1961possible, chalk2018toward}. Although many forms of compression operate by reducing redundancy, the notion of neural compression is contentious because compressive codes cannot by themselves explain observed differences in the representational capacity of lower-level visual regions versus higher-order cognitive regions \cite{barlow2001redundancy}. Thus, a gap exists in the evidence for neural compression occurring heterogeneously across lower-level to higher-order circuits at the macroscale.

Here we address this gap by testing a network theory of compression in a large sample of youth ($n=1,040$; ages 8 to 23 years). The theory formulates compression as a function of communication dynamics using random walks on networks. In a random walk, regions communicate with other regions by propagating signals called walkers---which represent repeated spatiotemporal motifs of activity---along structural connections to neighboring regions according to probabilities proportional to connection strengths \cite{avena2018communication}. The validity of the random walk model is supported by prior evidence that random walks on brain networks predict the strength and direction of activity \cite{goni2014resting, Seguin2019InferringNS, betzel2022multi}, the synaptic spread of pathogens \cite{Henderson2019SynucleinPS}, the evolutionary complexity of neuroanatomy across mammalian species \cite{griffa2022evolution}, and the speed and accuracy of human behavior \cite{zhou2022efficient}. We consider how random walks on networks behave according to rate-distortion functions, which are the basis of lossy compression, to predict different levels of compression in circuits \cite{Shannon1959CodingTF}.

To test the network theory of compression, we quantify the dimensionality and representational capacity of circuits. Low-level to higher-order circuits were sampled along a so-called \textit{principal gradient} of organization, anchored at one end by sensorimotor functions and at the other by higher-order cognitive functions \cite{Huntenburg2018LargeScaleGI}. We measure the dimensionality of spontaneous (rather than evoked) circuit activity because it combines internally generated information with accurate representations of sensory stimuli \cite{avitan2022not}. Measuring dimensionality is an appropriate test of the theory because dimensionality reduction is a form of compression \cite{Humphries2020StrongAW}. Due to information loss, compression also constrains the capacity to represent relevant information \cite{fusi2016neurons, chung2021neural}. Therefore, we measure the representational capacity of biological circuits and recurrent neural networks (RNNs) using automated meta-analysis and supervised learning for a wide range of behaviors \cite{yarkoni2011large, Yang2019TaskRI}. We find that circuits that support compression have low-dimensional activity and reduced representational capacity. 

\section{Method}

\textbf{Participants}. We used a subset ($n=1,040$) of an existing dataset with low in-scanner motion and high quality images \cite{satterthwaite2014neuroimaging}. Spontaneous activity was measured over 6-minute resting-state fMRI scans; motion artifact and other biological signals were removed. Structural networks were constructed from each subject's diffusion weighted imaging data  \cite{yeh2013deterministic}. Nodes of the network represented 360 cortical gray matter regions \cite{glasser2016multi}. Edges of the network represented the microstructure of white matter tracts as measured by fractional anisotropy. See the \textbf{Appendix} for more details and equations.

\begin{figure}[h]
	\centering
	\includegraphics[width=0.9 \columnwidth]{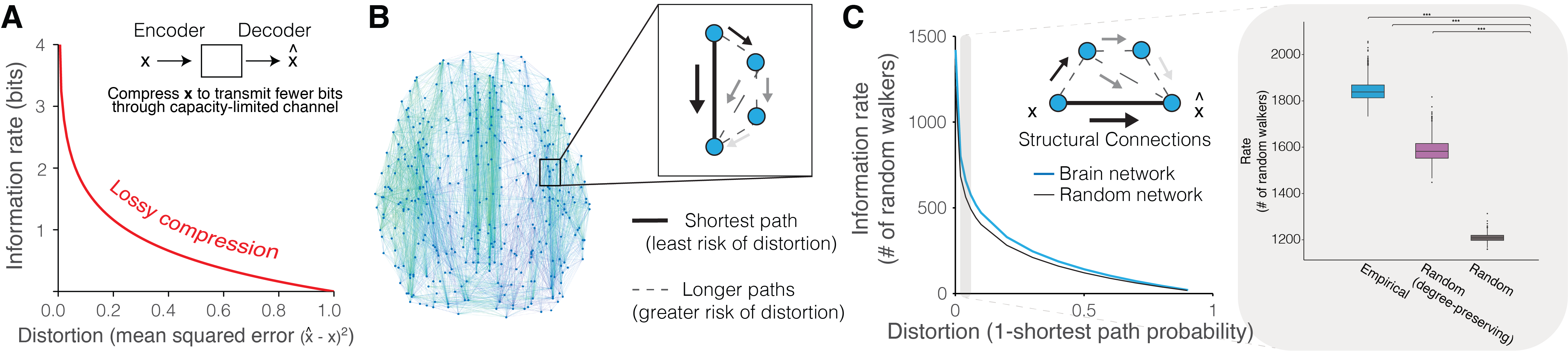}
	\caption{\textbf{Rate-distortion functions using random walks on networks.} \textbf{\emph{(A)}} Lossy compression improves information rate efficiency at the cost of distortion. \textbf{\emph{(B)}} In structural brain networks, we define the distortion as the probability of a random walker \textit{not} taking the shortest path because of noise introduced by longer paths. \textbf{\emph{(C)}} Rate-distortion functions for structural networks (blue) differ from those for random networks (black). Grey inset: at a distortion of 0.01, individual brain networks ($n=1,040$) compress less than expected, suggesting brain networks prioritize transmission fidelity.}
	\label{fig1}
\end{figure}

\textbf{Rate-distortion function}. We operationalized compression as the ability to send fewer random walkers across networks according to a rate-distortion function \textbf{(Figure \ref{fig1})} \cite{Shannon1959CodingTF}. Rate-distortion functions govern the fidelity and cost of lossy compression under an error-correcting code \cite{cover1991information}. A random walk model naturally implements a repetition code, whereby walkers are sent repeatedly to enhance resilience to errors. We operationalize the \textit{information rate} as the number of random walkers sent under a repetition code. The fidelity of information transfer would then depend on whether walkers could access shortest paths because visiting more regions risks corrupting the original signal. Thus, we operationalize the \textit{distortion} introduced by channel noise as the probability of a walker \textit{not} taking shortest paths across the structural network. The chance to take the shortest path can be increased by sending more walkers but this enlarges the transmission. When fewer walkers can achieve the same chance to take the shortest path, we operationalized fewer walkers as compression. We count the number walkers by dividing a given probability of walkers not taking shortest paths by the probability that walkers do not take shortest paths \cite{goni2013, fornito2016chapter}. Prior research supports the validity of our operationalization, which is consistent with hypotheses about how rate-distortion functions alter with system changes, with the cost of error, and in high- or low-fidelity regimes \cite{zhou2022efficient, sims2018efficient, marzen2017evolution}. 

\textbf{Dimensionality}. Dimensionality is commonly measured by the discrete number of components that are detected by eigendecomposition methods like principal components analysis and that explain a specified amount of variance. We use the \textit{participation ratio}, which is an eigenspectrum metric that provides a continuous value of dimensionality  \cite{gao2017theory, litwin2017optimal}. 

\textbf{Representational capacity}. We quantify the representation of 24 human behaviors across brain circuits using automated meta-analyses of neuroimaging literature \cite{yarkoni2011large}. This procedure provides statistical maps that associate regional activity with 24 terms that range from visual perception to emotion. Maps were validated against manual approaches and conventional maps of anatomical function. We measure representational capacity as the number of significantly associated terms. We also quantify representation of 20 multi-sensory working memory variables in RNNs trained to perform working memory tasks with supervised learning \cite{Yang2019TaskRI}. The recurrent matrix included 256 leaky units producing activity with a softplus activation function that mimics neuronal nonlinearity. We measure the representational capacity as the amount of unit activity variance across trials per task, which prior work used to quantify the amount of represented stimulus information.

\section{Results}

\begin{figure}[h]
	\centering
	\includegraphics[width=0.815 \columnwidth]{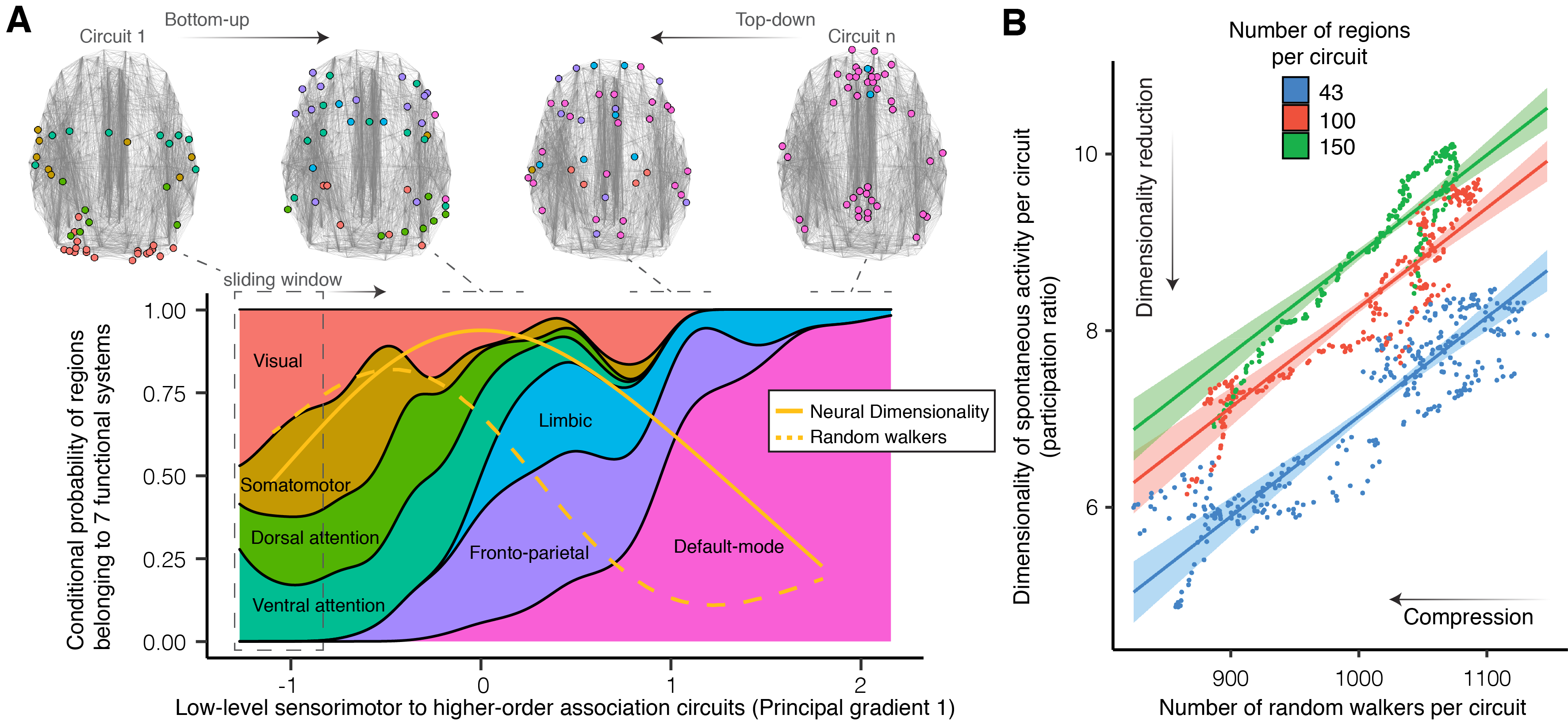}
	\caption{\textbf{Circuit compression explains the dimensionality of spontaneous activity.} \textbf{\textit{(A)}} Dimensionality and random walkers across circuits that are sampled along the principal gradient. For visual comparison, values are normalized between 0 and 1 and reported with the conditional probability of regions belonging to 7 functional systems. \textbf{\textit{(B)}} Circuit compression predicts the dimensionality of spontaneous activity across circuits of different sizes.}
	\label{fig2}
\end{figure}

\begin{figure}[h]
	\centering
	\includegraphics[width=0.7 \columnwidth]{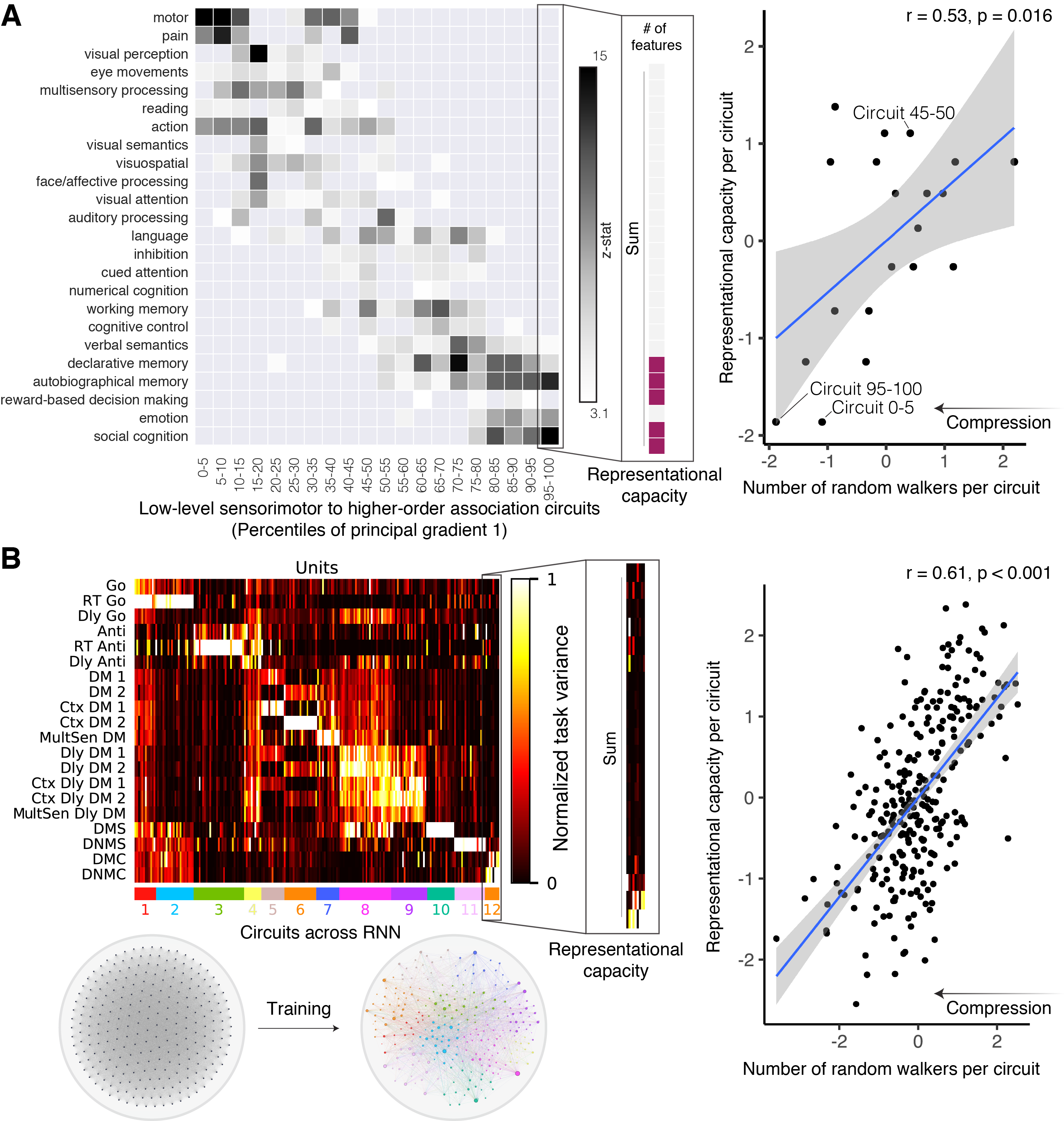}
	\caption{\textbf{Circuit compression explains the representational capacity of biological and artificial neural circuits.} \textbf{\emph{(A)}} \textit{Left}: Circuits along a sensorimotor to association continuum (columns) have varying representational capacity for information relevant to diverse behaviors (rows). \textit{Right}: Circuit compression constrains representational capacity. \textbf{\emph{(B)}} \textit{Left}: Example of one RNN with 256 units (columns) that was trained with supervised learning to perform 20 multisensory working memory tasks (rows; see \textbf{Appendix} for legend). Following training, RNNs learned to use specialized circuits depicted as 12 colored clusters from $t$-distributed stochastic neighbor embedding. Matrix entries show different representational capacity for each task defined by the task variance. \textit{Right}: Circuit compression constrains representational capacity. Matrices reproduced from code in \cite{margulies2016situating}  and \cite{Yang2019TaskRI}.
	\label{fig3}} 
\end{figure}

Circuit compression predicts low-dimensional spontaneous activity and reduced representational capacity. We slid a window of binned regions across the principal gradient to sample low-level to higher-order circuits. For each circuit, we measured compression and dimensionality (\textbf{Figure \ref{fig2}}). We tested linear and non-linear effects while minimizing over-fitting using generalized additive models, a non-parametric regression method. As predicted, circuits that compress information by using fewer random walkers had lower-dimensional activity (linear effect $t$=12.13, $p<$0.001). Compression and dimensionality are inverted-U functions along the principal gradient (non-linear effect $F$=350.9, $p$<0.001). As predicted, circuits with more compression had reduced representational capacity in humans ($r=0.53, p=0.016$) and in RNNs ($r=0.61$, $p<0.001$) for diverse behaviors (\textbf{Figure \ref{fig3}}).

\section{Discussion}

Consistent with the cognitive compression hypothesis, our findings link the function of higher-order cognitive circuits to compression and low-dimensional activity. We found that dimensionality and compression are related and are non-linear functions of circuit hierarchy: highest compression exists at the levels of sensorimotor and association function and lowest compression exists at the level of attention. Circuits with more compression represented fewer behaviors, explaining how circuits may efficiently specialize their function. Our findings relate information transmission to dimensionality, a property thought to support information processing. Expanding dimensionality improves representation and classification, whereas reducing dimensionality improves generalizability to unfamiliar contexts \cite{ganguli2012compressed, tang2019effective, sydnor2021neurodevelopment, jazayeri2021interpreting, chung2021neural}. Heterogenous levels of dimensionality support flexible behavior and compositionality of compact representations \cite{Cole2013MultitaskCR, ito2022constructing, ma2022}. To further understand information processing, the model could be used to explain how the brain generates and updates compressed models of the environment with differing constraints of dimensionality, timescales, and controllability \cite{Murray2014AHO, gu2015controllability, gao2017theory, jazayeri2021interpreting}, key elements of predictive coding \cite{clark2013whatever, himberger2018principles, chalk2018toward}. Our model admits several limitations. Although investment of metabolic resources and myelin improves a repetition code's efficiency, the code by itself is rudimentary and has limited efficiency, motivating studies of more sophisticated compression \cite{zhou2022efficient}. Strengths of the model include being robust at the individual level and bridging the compression of brain regions to behavior. Hence, future work could investigate if skewed levels of compression explain how circuit computations go awry in mental illnesses due to over-compression that results in inaccurate and highly generalized models of the world. 

%%%%%%%%%%%%%%%%%%%%%%%%%%%%%%%%%%%%%%%%%%%%%%%%%%%%%%%%%%%%

\section{Acknowledgments}

D.Z. acknowledges support from the National Institute of Mental Health (F31MH126569). D.S.B. acknowledges support from the John D. and Catherine T. MacArthur Foundation, the Swartz Foundation, the Paul G. Allen Family Foundation, the Alfred P. Sloan Foundation and the NSF (IIS-1926757). D.S.B. and T.D.S. acknowledge support from the National Institute of Mental Health (R01MH113550).

\appendix

\section{Appendix}

\subsection{Participants}
Participants completed a cross-sectional imaging protocol. Exclusion criteria included excessive in-scanner motion, serious health conditions, and quality control models. The participants used in this paper are a subset of the 1,601 participants who completed the cross-sectional imaging protocol. We excluded participants with health-related exclusionary criteria (n = 154) and with scans that failed a rigorous quality assurance protocol for diffusion-weighted imaging (DWI; n = 162). We further excluded subjects with incomplete or poor ASL and field map scans (n = 60). Finally, participants with poor quality T1-weighted anatomical reconstructions (n = 10) were removed from the sample. The final sample contained 1042 subjects (mean age = 15.35, SD = 3.38 years; 467 males, 575 females). All adult participants provided informed consent; all minors provided assent and their parent or guardian provided informed consent.

\subsection{Formulating rate-distortion functions using random walks on networks}

Our formulation of the rate-distortion function uses exact expressions for the mean first passage time of random walks on networks \cite{noh2004random}. We operationalize rate-distortion functions as the number of random walkers $r_{i j}$ which begin at node $i$ that were required for at least one to travel along the shortest path to another node $j$ with probability $\eta$ \cite{goni2013, fornito2016chapter}. We define the transition probability matrix as $\mathbf{U}=\mathbf{W} \mathbf{L}^{-1}$, where each entry $W_{i j}$ describes the weight of the connection from node $i$ to node $j$, and each entry of the diagonal matrix $\mathbf{L}$ is the strength of each node $i$ given by $\sum_i W_{i j}$. Intuitively, each entry $U_{i j}$ defines the probability of a random walker traveling from node $i$ to node $j$ in one step. 

We operationalized distortion as the probability that one of the sent walkers does \textit{not} take the shortest path. To obtain this probability we first need to compute the probability that a random walker travels from node $i$ to node $j$ along the shortest path. We define a new matrix $U'(i)$ that is equivalent to $\mathbf{U}$ but with the non-diagonal elements of row $i$ set to zero and $U_{ii}=1$ as an absorbant state. Then, the probability of randomly walking from $i$ to $j$ along the shortest path is given by $1 - \sum_{n = 1}^N [U'(i)^H]_{in}$, where $H$ is the number of connections composing the shortest path from $i$ to $j$. 

We can use these definitions to determine the number of random walkers $r$ required to guarantee (with probability $\eta$) that at least one of them travels from $i$ to $j$ along the shortest path: 

\begin{equation}
r_{i j}(\eta)=\frac{\log (1-\eta)}{\log \left(\sum_{n=1}^{N}\left[U'(i)^{H}\right]_{in}\right)}.
\end{equation}
In our analyses, we calculate $r_{i j}$ for $\eta$ $= 0.999$ for each participant. This calculation returns a right stochastic matrix $U'_{i}$, where the number of random walkers a brain region receives is $(r_{j i}(\eta))$ averaged over $j$. 

Finally, we substitute these definitions as our choice of the rate-distortion function. A signal $x$ is encoded as $\hat{x}$ with a level of distortion $D$ that depends on the information rate $R$. We define the distortion function of any signal $x$ from brain region $i$ to a compressed representation $\hat{x}$ decoded in brain region $j$ as $d(x, \hat{x})_{i j}=(1-\eta)$, where $\eta$ denotes the probability that a walker gets from node $i$ to node $j$ along the shortest path. The rate-distortion function is $R(D) \equiv r_{i j}(d(x, \hat{x})_{i j})$, where the rate is $r_{i j}(\eta)$ required to achieve a tolerated level of distortion $d(x, \hat{x})_{i j}$.

\subsection{Dimensionality}

Principal components analysis is a common multivariate method to decompose data into a mutually orthogonal (uncorrelated) set of data ordered by the amount of variance explained \cite{Pang2016DimensionalityRI, gotts2020brain}. In principal components analysis, the principal components are the eigenvectors $v$, and the proportion of variance explained by the components are the eigenvalues $\lambda$ of the demeaned and centered $N \times N$ covariance matrix of spontaneous activity. The participation ratio normalizes eigenvalues $\tilde{\lambda_i}$ with the equation: $\tilde{\lambda_i} = \frac{\lambda_i}{\sum_j \lambda_j}$ \cite{gao2017theory, litwin2017optimal}. The dimensionality then is: 

\begin{equation}
\frac{1}{\sum_i \tilde{\lambda_i}^2}.
\end{equation}

\subsection{Imaging acquisition}

Diffusion imaging data and all other MRI data were acquired on the same 3T Siemens Tim Trio whole-body scanner and 32-channel head coil at the Hospital of the University of Pennsylvania. DWI scans were obtained using a twice-focused spin-echo (TRSE) single-shot EPI sequence (TR = 8100 ms, TE = 82 ms, FOV = 240 mm$^2$/240 mm$^2$; Matrix = RL: 128/AP:128/Slices:70, in-plane resolution (x \& y) 1.875 mm$^2$; slice thickness = 2 mm, gap = 0; FlipAngle = 90°/180°/180°, volumes = 71, GRAPPA factor = 3, bandwidth = 2170 Hz/pixel, PE direction = AP). The sequence employs a four-lobed diffusion encoding gradient scheme combined with a 90-180-180 spin-echo sequence designed to minimize eddy current artifacts. The complete sequence consisted of 64 diffusion-weighted directions with $b$ = 1000 s/mm$^2$ and 7 interspersed scans where $b$ = 0 s/mm$^2$. Scan time was about 11 min. The imaging volume was prescribed in axial orientation covering the entire cerebrum with the topmost slice just superior to the apex of the brain \cite{roalf2016impact}.
	
\subsection{Connectome construction}

Cortical gray matter was parcellated according to the Glasser atlas \cite{glasser2016multi}, defining 360 brain regions as nodes for each subject's structural brain network, denoted as the weighted adjacency matrix $\mathbf{A}$. DWI data was imported into DSI Studio software and the diffusion tensor was estimated at each voxel \cite{yeh2013deterministic}. For deterministic tractography, whole-brain fiber tracking was implemented for each subject in DSI Studio using a modified fiber assessment by continuous tracking (FACT) algorithm with Euler interpolation, initiating 1,000,000 streamlines after removing all streamlines with length less than 10mm or greater than 400mm. Fiber tracking was performed with an angular threshold of 45, a step size of 0.9375mm, and a fractional anisotropy (FA) threshold determined empirically by Otzu’s method, which optimizes the contrast between foreground and background \cite{yeh2013deterministic}. FA was calculated along the path of each reconstructed streamline. For each subject, edges of the structural network were defined where at least one streamline connected a pair of nodes. Edge weights were defined by the average FA along streamlines connecting any pair of nodes. The resulting structural connectivity matrices were not thresholded and contain edges weighted between 0 and 1.

\subsection{Metabolic energy}

\begin{figure}[htbp!]
	\centering
	\includegraphics[width=0.3 \columnwidth]{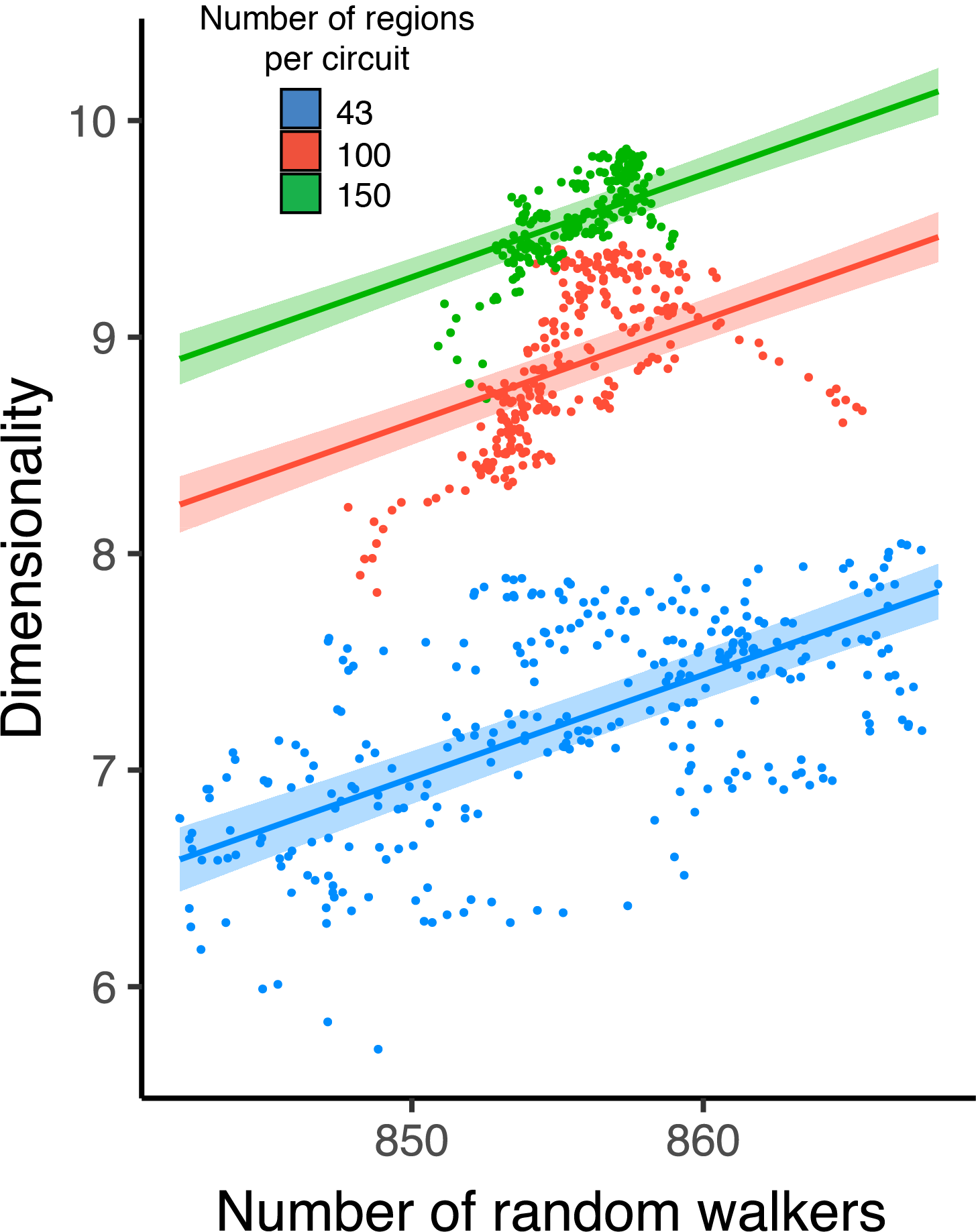}
	\caption{\textbf{Circuit compression explains the dimensionality of spontaneous activity.} \textbf{\textit{(A)}} The number of random walkers (unnormalized for CBF) predicts the dimensionality of spontaneous activity across circuits of different sizes. Plotted with a generalized additive model, which is a non-parametric regression method.
	\label{fig2_supplement}} 
\end{figure}

Compressed codes economically use limited metabolic resources, which is a key benefit of efficient coding \cite{laughlin2001energy}. Metabolic energy expenditure is tightly linked to regional cerebral blood flow \cite{raichle2006brain, lecrux2019reliable}. Therefore we normalize our predictions of compression by dividing the number of random walkers by the regional cerebral blood flow (CBF), an operationalization of metabolic energy expenditure \cite{gur2009regional, vaishnavi2010regional}. Arterial-spin labeling data was used to determine the regional cerebral blood flow (mL per 100 g tissue per second). We use this normalized metric when predicting the dimensionality of spontaneous activity. The normalized metric is in units of number of random walkers per mL per 100 g tissue per second. Sensitivity analysis suggests that our theory remains predictive of dimensionality without this normalization step (\textbf{Figure \ref{fig2_supplement}}). Circuits that transmit fewer random walkers to the whole brain have low-dimensional spontaneous activity.

CBF was quantified from control-label pairs using ASLtbx \cite{wang2008empirical}, as previously described \cite{satterthwaite2014impact}. We consider $f$ as CBF, $\delta M$ as the difference of the signal between the control and label acquisitions, $R_{1a}$ as the longitudinal relaxation rate of blood, $\tau$ as the labeling time, $\omega$ as the post-labeling delay time, $\alpha$ as the labeling efficiency, $\lambda$ as the blood/tissue water partition coefficient, and $M_0$ as the approximated control image intensity. Together, CBF $f$ can be calculated according to the equation:

\begin{equation}
f=\frac{\Delta M \lambda R_{1 a} \exp \left(\omega R_{1 a}\right)}{2 M_{0} \alpha}\left[1-\exp \left(-\tau R_{1 a}\right)\right]^{-1}. 
\end{equation}

Because prior work has shown that the T1 relaxation time changes substantially in development and varies by sex, this parameter was set according to previously established methods, which enhance CBF estimation accuracy and reliability in pediatric populations \cite{wu2010vivo, jain2012longitudinal}.

\subsection{Principle gradient 1}

The principal gradient of intrinsic functional connectivity reflects a functional processing hierarchy from sensory cortex to association cortex. We used both a publicly available atlas of the principal gradient and published tools to calculate the principal gradient in our own data \cite{vos2020brainspace}. Briefly, these tools use a form of Lacplacian eigenmapping, a non-linear dimensionality reduction method, on the intrinsic functional connectivity data to find latent dimensions in the data. For each participant, we identified the latent dimension that was most similar to the published atlas of the principal gradient \cite{margulies2016situating}. We then averaged the latent dimension across individuals to obtain regional values. The principal gradient in our data was highly correlated with the previously published atlas ($r$=0.90) as well as to the average anterior-to-posterior anatomical coordinates ($r$=0.91).

\subsection{Representational capacity and Neurosynth}

Neurosynth is an automated meta-analysis tool that uses natural language processing to search for the co-occurrence of behavioral terms and fMRI activation in a standardized coordinate form across over 10,000 studies \cite{yarkoni2011large}. The tool statistically links cognitive processes and mental states to regional brain activity. For replicability, we used previously described methods, the published principal gradient, and the version of Neurosynth accessed at the time of the previous study to create masks for the regions of interest \cite{margulies2016situating}. Briefly, the published atlas was projected to a 2-mm volumetric space for our atlas (\href{https://github.com/PennLINC/xcpEngine/tree/master/atlas/glasser360}{available here}). The volumetric map was binned into five-percentile increments and binarized. Each of these 20 maps ranging from 0–5\% to 95–100\% were used as inputs to the meta-analysis. For each map, the output of the meta-analysis was a $z$-statistic associated with the feature term. The feature terms were derived from the set of 50 topic terms (\href{https://github.com/NeuroanatomyAndConnectivity/}{as in previous work}. This previous work removed 30 terms because they were above the statistical association threshold of z$>$3.1, removed 6 as presumed noise terms because they were not consistent with any particular cognitive function. The final set consisted of 24 distinct classes of behavior. For visualization, the terms were then ordered according to the weighted mean of the feature association.

\subsection{RNNs and task variance}

\begin{figure}[htbp!]
	\centering
	\includegraphics[width=0.6 \columnwidth]{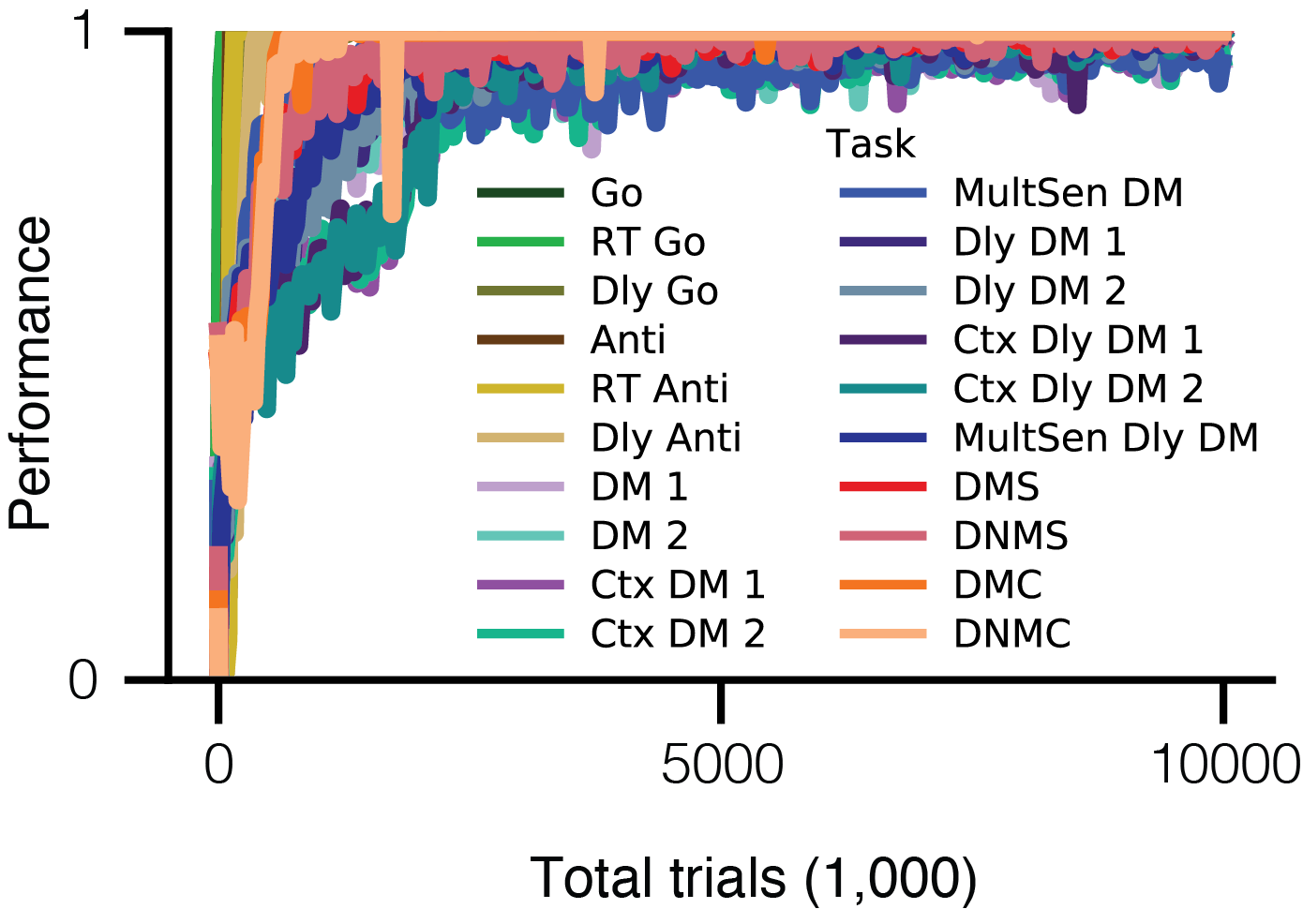}
	\caption{\textbf{Trained RNNs effectively perform working memory tasks.} Task performance increases with training. Reproduced with code from \cite{Yang2019TaskRI}.
	\label{fig1_supplement}} 
\end{figure}

Following \cite{Yang2019TaskRI}, we reproduced figures showing that the same RNN can effectively perform multiple working memory tasks following training (\textbf{Figure \ref{fig1_supplement}}). In this work, we analyzed the recurrent matrix, and for replicability we used the 20 trained networks from the prior study. The recurrent matrix consisted of 256 units with positive (excitatory) and negative (inhibitory) connections; across all 20 networks, these connection strengths range from $-4.27$ to $2.99$. Noisy input units encoded task variables and output motor units read out nonlinearly to encode actions, such as a saccade or reach direction. A supervised learning protocol was used to modify all connection weights (input, recurrent, and output) to minimize the difference between the output units and the correct task output. 

To match the range of values in the connectivity matrices extracted from human brain networks, we scaled the values of the recurrent connection matrix to between 0 and 1. When modeling random walks on these networks, this scaling also serves to bias the behavior of random walkers that represent activity flow to mimic inhibitory signaling. Specifically, connections that were originally more negative become lower probability nodes which inhibit the subsequent propagation to neighboring pathways and the overall flow of activity to inhibited units. Inhibitory pathways thus decrease the number of random walkers and redirect them away from locally connected, similar circuits (or clusters) of units towards different excitatory pathways.

The RNNs were trained to perform 20 tasks. Tasks included the Go, Reaction-time go (RT Go), Delayed go (Dly Go), Anti-response (Anti), Reaction-time anti-response (RT Anti), Delayed anti-response (Dly Anti), Decision making 1 (DM 1), Decision making 2 (DM 2), Context-dependent decision making 1 (Ctx DM 1), Context-dependent decision making 2 (Ctx DM 2), Multi-sensory decision making (MultSen DM), Delayed decision making 1 (Dly DM 1), Delayed decision making 2 (Dly DM 2), Context-dependent delayed decision making 1 (Ctx Dly DM 1), Context-dependent delayed decision making 2 (Ctx Dly DM 2), Multi-sensory delayed decision making (MultSen Dly DM), Delayed match-to-sample (DMS), Delayed non-match-to-sample (DNMS), Delayed match-to-category (DMC), and Delayed non-match-to-category (DNMC). These tasks were simplified from tasks used in prior non-human animal experiments and computational studies \cite{funahashi1989mnemonic, munoz2004look, gold2007neural, mante2013context, raposo2014category, romo1999neuronal, miller1996neural, freedman2016neuronal}. Collectively, these tasks involve cognitive processes including pro-response, working memory, anti-response, decision making, gating modulation, integrating modulation, comparison, and categorization.

The task variance $\mathrm{TV}_i(A)$ for task $A$ and unit $i$ was calculated over many trials for each RNN unit. The averaged variance across all trials is the task variance:

\begin{equation}
\mathrm{TV}_i(A)=\left\langle\left[f_i(j, t)-\left\langle f_i\left(j^{\prime}, t\right)\right\rangle_{j^{\prime}}\right]^2\right\rangle_{j, t}
\end{equation}

where $f_i(j, t)$ is the activity (non-negative and non-saturating firing rate) of unit $i$ on time $t$ of trial $j$.

\subsection{Statistical analysis}

We used generalized additive models with penalized splines, a non-parametric regression method, allowing for statistically rigorous modeling of linear and non-linear effects while minimizing over-fitting \cite{wood2004stable}. To predict dimensionality we included covariates for the number of random walkers, the average degree of connectivity, window size, and splines for the principal gradient ($k=4$) and its interaction with window size. To assess the relationship between the number of random walkers and the representational capacity, we used a Pearson correlation coefficient. 

\newpage 
\bibliography{./bibliography}

\end{document}